# Consistent analytical solution of the time-dependent Schrödinger equation for nanoscale circuits with laser-assisted quantum tunneling

Mark J. Hagmann and Logan D. Gibb
NewPath Research LLC, 2880 s. Main Street, Suite 214, Salt Lake City, Utah, USA 84115
(Dated June 8, 2020)

**ABSTRACT**

It is now common practice to solve the Schrödinger equation to estimate the tunneling current between two metal electrodes at specified potentials, or the transmission through a potential barrier by assuming an incident, reflected, and transmitted wave. However, we suggest that these methods may not be appropriate for nanoscale circuits. The electron man-free path may be as long as 68.2 nm in metallic elements so we consider the possibility that quantum effects may occur throughout a nanoscale circuit, including the connections. Analytical methods are presented for modeling the coherent transfer of the wavefunction through a closed circuit.

## I. INTRODUCTION

In 1991 Kalotas and Lee [1] introduced the transfer-matrix method for solving the one-dimensional Schrödinger equation to model quantum tunneling in arbitrary static potential barriers. Later Grossel, Vigoureux and Baida [2] compared the stability of this method with WKB. We were the first to apply the transfer-matrix method with a time-dependent potential in studying the effects of the barrier traversal time in laser-assisted scanning tunneling microscopy (STM) [3]. These simulations guided our development of microwave oscillators based on laser-assisted field emission [4] and the generation of microwave frequency combs by focusing a mode-locked laser on the tunneling junction of an STM [5]. Now we are studying a variant of STM where extremely low-power ($\approx$ 3 atto-watt) microwave harmonics of the laser pulse repetition rate have a signal-to-noise ratio of 20 dB. This extremely low-noise measurement is made possible because the quality factor (Q) at each harmonic is approximately $10^{12}$, which is five times that for a cryogenic microwave cavity [6]. Thus, the effect of white noise on the measurements is greatly reduced. These harmonics can provide faster and more stable feedback control of the tip-sample separation without exposing the sample to the continuous intense static field of $\approx 10^9$ V/m in an STM [7].

## II. DELIMITATIONS

We model electrons tunneling between two metal electrodes with the same work function and zero resistivity. The distribution of electron energies in these electrodes and the effects of images of the tunneling electrons at the two electrodes are neglected to obtain relatively-simple analytical solutions. The Dirac equation should be used to include the properties of the electron (Per-Olov Löwdin, personal communication, 1998) but we still follow the convention of using the Schrödinger equation. Only metal electrodes are considered in our simulations without addressing the possible use of semiconductors or superconductors.

## III. ANALYSIS OF RELEVANT WORK BY TIEN AND GORDON AS BACKGROUND

We begin with a brief description and analysis of the pioneering work by Tien and Gordon to place our work in context. In 1963 Tien and Gordon [8] published what appears to be the first analytical solution of the time-dependent Schrödinger equation for quantum tunneling in a sinusoidally-modulated triangular potential barrier. Google Scholar lists 996 papers referring to this paper in studies of quantum tunneling for a variety of nanostructures.

Previously we developed numerical methods to solve the Schrödinger equation by back-propagation in an arbitrary time-dependent potential [3] In that method a value is assumed for the coefficient in the transmitted wave. Then the coefficients are determined by stepping backward from segment to segment, and normalized to have a unit incident wave at the left-hand side. However, now we consider analytical solutions for circuit models related to Fig. 1 which is consistent with that used by Tien and Gordon. An ideal DC voltage source and an ideal AC voltage source are in series with the tunneling junction. The work function is effectively added as a series element in this model. The planar structure assumed for the tunneling junction is shown by the dashed lines in the figure. The letters "A" and "B" denote the two sides of the model. The coordinate x is zero at the surface of the cathode and equals "a" at the surface of the anode.

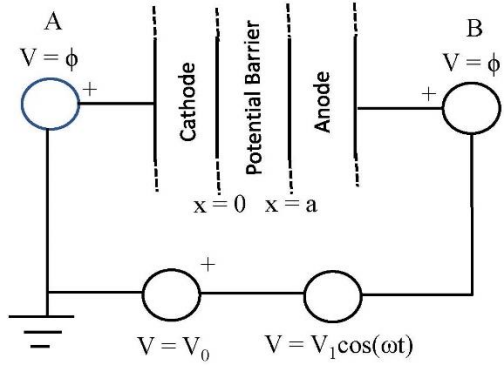

Fig. 1. Circuit model for the dynamic solution.

**Extension of the work by Tien and Gordon by introducing Airy functions**

Tien and Gordon [8] considered quantum tunneling between two electrodes when an applied DC electric field causes the wavefunction $\psi_0$ (x,y,z) shown in Eq. (1). They used the properties of the Hamiltonian to determine the change of this wavefunction when one electrode is grounded and the second has an applied sinusoidal potential of $\bar{V}\cos(\omega t)$ superimposed on the DC potential. They showed that, for any applied DC field, the wave function is modified as shown in Eq. (2)

$$\Psi_0 = \psi_0(x,y,z) e^{-i\frac{Et}{\hbar}} \qquad (1)$$

$$\Psi(x,y,z,t) = \Psi_0 \sum_{n=-\infty}^{n=+\infty} J_n\left(\frac{e\bar{V}}{\hbar\omega}\right) e^{-in\omega t} \qquad (2)$$

In Section 1 of the appendix it is shown that the solution for the wavefunction within a static barrier having a linear potential is given by Eq. (3) where Ai and Bi are Airy functions. The equations for the real constants A and B and the complex coefficients $C_1$ and $C_2$, are derived in that section. Using Eq. (4) with Eq. (A2-8) that is derived in Section 2 of the appendix we obtain Eq. (5) which is more convenient to use when satisfying the boundary conditions.

$$\psi(x) = C_1 Ai\left(\frac{B-x}{A}\right) + C_2 Bi\left(\frac{B-x}{A}\right) \qquad (3)$$

$$\Psi(x,t) = \left[C_1 Ai\left(\frac{B-x}{A}\right) + C_2 Bi\left(\frac{B-x}{A}\right)\right] e^{-i\frac{Et}{\hbar} - i\frac{e\bar{V}}{\hbar\omega}\sin(\omega t)} \qquad (4)$$

$$\Psi(x,t) = \left[C_1 Ai\left(\frac{B-x}{A}\right) + C_2 Bi\left(\frac{B-x}{A}\right)\right] e^{-i\frac{Et}{\hbar}} \sum_{n=-\infty}^{n=+\infty} J_n\left(\frac{e\bar{V}}{\hbar\omega}\right) e^{-in\omega t} \qquad (5)$$

Equation (6) is a general expression for the current density. Substituting Eq. (7), as the complete time-dependent wavefunction, into Eq. (6) gives Eq. (8) which has no time-dependence and is also independent of the values of $\bar{V}$ and $\omega$.

$$J_X(x,t) = \frac{-ie\hbar}{2m}\left(\Psi\frac{d\Psi^*}{dx} - \Psi^*\frac{d\Psi}{dx}\right) \qquad (6)$$

$$\Psi(x,t) = \psi(x) e^{-i\frac{Et}{\hbar}} e^{-i\frac{e\bar{V}\sin(\omega t)}{\hbar\omega}} \qquad (7)$$

$$J_X(x) = \frac{-ie\hbar}{2m}\left[\psi\frac{d\psi^*}{dx} - \psi^*\frac{d\psi}{dx}\right] \qquad (8)$$

Note that the current density in Eq. (8) is exactly that for the static case. This may be understood because, as Tien and Gordon noted [8], the time-dependent voltage does not change the spatial distribution of the wave function, but can only adiabatically modify the electron energies. Thus, the altered spectrum of energies may change the current in a superconductor but not in metal electrodes.

We acknowledge that Tien and Gordon presented a valid solution as a special case for the time-dependent Schrödinger equation but we question the many applications in which this analysis has been used that do not pertain to superconductors. Furthermore, we do not consider their solution to be unique. Other analyses that are also based on the transfer Hamiltonian formulation have shown that metal-barrier-metal (MBM) diodes and uncooled Schottky diodes are unlikely to find significant applications as mixers because superconducting Schottky diodes have a response with considerably greater nonlinearity [9]. We find this conclusion to be consistent with the analysis by Tien and Gordon, and again, to be clear, the approach of Tien and Gordon do not predict that adding a time-dependent potential would cause any change in the measured current when using metal electrodes.

**Derivation of the potential energy that is implicit in the analysis by Tien and Gordon**

Tien and Gordon [8] used the properties of the Hamiltonian to determine the change in the wavefunction for quantum tunneling when a sinusoidal potential is superimposed on the DC potential applied between two electrodes. Thus, there are both static and sinusoidal uniform electric fields in the gap. Tien and Gordon showed that the wavefunction is given by Eq. (9) where the static solution for $V_1 = 0$ is given in Eq. (10).

$$\Psi(x,y,z,t) = \Psi_0 \sum_{n=-\infty}^{n=+\infty} J_n\left(\frac{eV_1}{\hbar\omega}\right) e^{-in\omega t} \qquad (9)$$

$$\Psi_0 = \psi(x,y,z) e^{-i\frac{Et}{\hbar}} \qquad (10)$$

Equation (11) is an identity derived in Section 2 of the appendix by combining two equations labeled 8.514.5 and 8.515.6 in [10]. This identity has been confirmed by numerical testing. Using this identity with Eq. (9) gives Eq. (12) as the wavefunction.

$$e^{-i\alpha \sin(\beta)} \equiv \sum_{n=-\infty}^{n=+\infty} J_n(\alpha) e^{-in\beta} \tag{11}$$

$$\Psi(x,y,z,t) = \Psi_0 e^{-i\frac{eV_1 \sin(\omega t)}{\hbar \omega}} \tag{12}$$

Consider the time-dependent Schrödinger equation for a one-dimensional potential barrier V(x,t) in Eq. (13). Now Eq. (14), which is formed by rearranging Eq. (13), is used to determine the potential barrier corresponding to the wavefunction derived by Tien and Gordon.

$$\frac{\hbar^2}{2m} \frac{\partial^2 \Psi}{\partial x^2} - V(x,t)\Psi = -i\hbar \frac{\partial \Psi}{\partial t} \tag{13}$$

$$V(x,t) = \frac{i\hbar}{\Psi} \frac{\partial \Psi}{\partial t} + \frac{\hbar^2}{2m} \frac{1}{\Psi} \frac{\partial^2 \Psi}{\partial x^2} \tag{14}$$

The complete wavefunction given by Tien and Gordon for one spatial dimension may be written as Eq. (15).

$$\Psi(x,t) = \psi(x) e^{-i\left[\frac{Et}{\hbar} + \frac{eV_1 \sin(\omega t)}{\hbar \omega}\right]} \tag{15}$$

Substituting Eq. (15) into Eq. (14) gives the following expression for the potential barrier.

$$V(x,t) = E + eV_1 \cos(\omega t) + \frac{\hbar^2}{2m\psi} \frac{d^2 \psi}{dx^2} \tag{16}$$

Section 1 of the appendix is a derivation of the wavefunction in terms of Airy functions added for clarification because we have not seen a detailed presentation of this problem presented by others. Equations (17) and (18) are equivalent to Eqs. (A1-1-13) and (A1-8) in the appendix, and Eq. (19) is obtained by combining them to eliminate the independent variable ξ.

$$\frac{d^2 \psi}{d\xi^2} = -\xi \psi \tag{17}$$

$$\xi = \frac{x - B}{A} \tag{18}$$

$$\frac{d^2 \psi}{dx^2} = \frac{(B-x)}{A^3} \psi \tag{19}$$

Substituting the expression for the second derivative in Eq. (19) into Eq. (16) gives Eq. (20). Then substituting the definition for A from Eq. (A1-11) from Section 1 of the appendix gives Eq. (21). Substituting the definition for B in Eq. (A1-12) in the Section 1 gives Eq. (22). This is simplified to Eq. (23) as the first expression that has been presented for the potential barrier which is implicit in the solution by Tien and Gordon to help in understanding and implementing their derivation.

$$V(x,t) = E + eV_1 \cos(\omega t) + \frac{\hbar^2}{2m} \frac{(B_1 - x)}{A^3} \tag{20}$$

$$V(x,t) = E + eV_1 \cos(\omega t) + (B_1 - x)\frac{V_0}{a} \tag{21}$$

$$V(x,t) = E + eV_1 \cos(\omega t) + \left[\left(1 + \frac{\phi - E}{V_0}\right) - \frac{x}{a}\right]V_0 \quad (22)$$

$$V(x,t) = \phi + \left(1 - \frac{x}{a}\right)V_0 + eV_1 \cos(\omega t) \quad (23)$$

## IV. CRITERIA FOR THE CONSISTENT SIMULATION OF A NANOSCALE CIRCUIT

### 1. Coherent propagation of the wavefunction

Others have presented solutions of the Schrödinger equation with three sections at constant potentials where there is an incident and a reflected wave at one end and a transmitted wave at the but this would not be consistent with a nanoscale device (e.g. [11], [12], [13], [14], [15]). This approach has pedagogical value, but does not consider how the incident, reflected, and transmitted waves may interact at the voltage source ("battery") that is connected to this system. In modeling a scanning tunneling microscope generally quantum effects are only considered in the region between the two electrodes, which may be ideal metals [16]. Now we consider the possibility that quantum effects may occur throughout the nanoscale circuit, including the connections, because the electron mean-free path may be as long as 68.2 nm in metallic elements [17]. We acknowledge that, surprisingly, the effective resistance of a nanoscale wire is proportional to the mean free path of the electrons [17]. The basis for this increase in resistance has been presented and discussed elsewhere [18], [19]. However, the path lengths are quite short which would limit the resistance. Furthermore, we are interested in nanoscales devices that are based on quantum tunneling, and this effect already introduces a high electrical resistance into the circuit. Thus, it appears that the effects of the increased resistance in the circuit may not be problematic in our application.

### 2. Circuit models for resistive loads and voltage sources

In quantum simulations generally a voltage source is not shown in the diagram but is implied by specifying the potential at two points. Now, in allowing for the coherent transfer of the wavefunction within a nanoscale circuit, a voltage source could be represented by a jump in the potential at a specific location or a linear rise in the potential over a specified distance. Figure 2 shows how a linear variation of the voltage may be presented as a linear variation in the potential for a voltage source and a load resistor having finite lengths with lossless connections. Note that the energy is greater than the potential to avoid quantum tunneling. In a one-dimensional static problem, the electrical current density in the x-direction is given by Eq. (24) as the product of the probability current density and the electron charge. Thus, the effective value of the resistance for the load resistor may be determined by dividing the voltage drop across the simulated load resistor by the product of the electrical current density and the effective cross-sectional area of the resistor. It would also be possible to model the resistor as a lumped circuit element by having a sharp drop in the potential as we generally model a voltage source by a jump in the potential but it could also be tapered as shown in Fig. 2. Adding resistors to a model is the converse to adding voltage sources. Horizontal lines, corresponding to a constant potential, represent wires that have no significant resistivity.

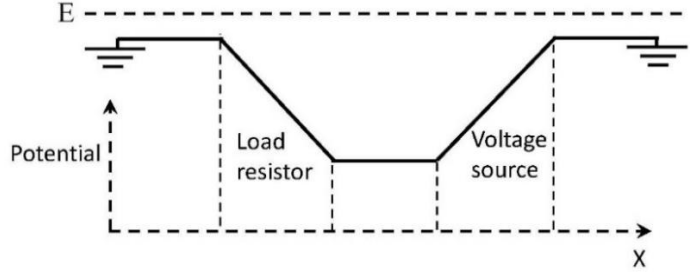

Fig. 2: Symmetric model for a load resistor, a voltage source, and lossless connectors.

$$J_X(x) = \frac{-ie\hbar}{2m}\left(\psi\frac{d\psi^*}{dx} - \psi^*\frac{d\psi}{dx}\right) \quad (24)$$

Figure 3 is a simple and consistent closed-loop model of a nanoscale circuit that has a load resistance as well as a tunneling junction. While linear or jump models are not shown for either the load resistor or the voltage sources, it is possible to approach the problem in the following manner: First, the effective voltage is specified which is defined as the voltage from the voltage source minus the voltage drop on the load resistor. The parameters $\phi$, $U_0$, a, E, and S are also specified so the wavefunction may be determined. Then the electrical current density is determined using Eq. (24). An effective cross-sectional area for the resistor may be input in order to provide the user with an estimate of the actual resistance. Then the voltage drop on the load resistor may be calculated so that the voltage of the voltage source may be determined to complete the solution. As in Fig. 2, we require that $0 < E < \phi - U_0$ so an electron will tunnel within the full length of the barrier.

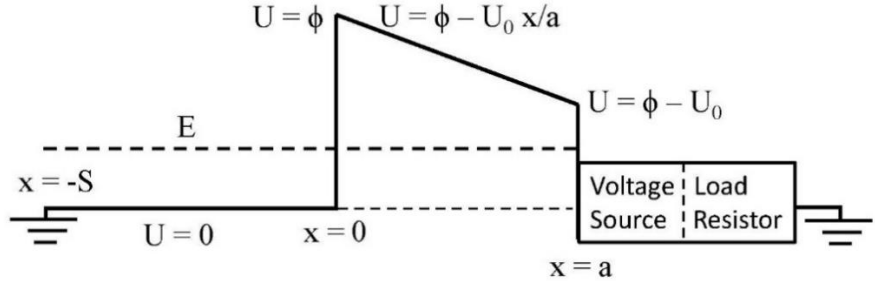

Fig. 3: Model for a closed-loop solution using a tunneling junction with a load resistor.

## V. EXAMPLE OF A CONSISTENT SOLUTION FOR A NANOSCALE CIRCUIT

Consider the model in Fig. 4 where Region 2 is the tunneling junction and Region 3 is the battery.

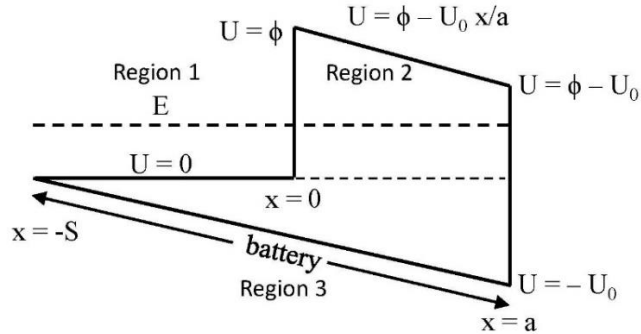

Fig. 4. Potential energy for second model of quantum tunneling in a static potential barrier.

With one spatial dimension x, the time-dependent Schrödinger equation is given by Eq. (25). With a static potential this simplifies to Eq. (26) where the full wavefunction is given as Eq. (27).

$$\frac{\hbar^2}{2m}\frac{\partial^2 \Psi}{\partial x^2} - U(x,t)\Psi = -i\hbar\frac{\partial \Psi}{\partial t} \quad (25)$$

$$\frac{\hbar^2}{2m}\frac{d^2 \psi}{dx^2} + [E - U(x)]\psi(x) = 0 \quad (26)$$

$$\Psi(x,t) = \psi(x)e^{-i\frac{Et}{\hbar}} \quad (27)$$

In Region 3, representing the battery where -S < x < a, the potential is given by Eq. (28). We substitute this into the Schrödinger equation Eq. (26) to obtain Eq. (29).

$$U_3(x) = -U_0\frac{(S+x)}{(S+a)} \quad (28)$$

$$\frac{\hbar^2}{2m}\frac{d^2\psi_3}{dx^2} + \left[E + \frac{U_0 S}{(S+a)} + \frac{U_0 x}{(S+a)}\right]\psi_3 = 0 \quad (29)$$

A change of variables shown in Eq. (30) is used with Eq. (29) to obtain Eq. (31) where the coefficients $A_3$ and $B_3$ have units of meters. Then Eq. (31) is rearranged to obtain Eq. (32).

$$x = A_3\xi + B_3 \quad (30)$$

$$\frac{\hbar^2}{2mA_3^2}\frac{d^2\psi_3}{d\xi^2} + \left[E + \frac{U_0 S}{(S+a)} + \frac{U_0(A_3\xi + B_3)}{(S+a)}\right]\psi_3 = 0 \quad (31)$$

$$\frac{d^2\psi_3}{d\xi^2} + \frac{2mA_3^2}{\hbar^2}\left[E + \frac{U_0(S+B_3)}{(S+a)}\right]\psi_3 + \frac{2mA_3^3 U_0}{\hbar^2(S+a)}\xi\psi_3 = 0 \quad (32)$$

Parameter $A_3$ is chosen by setting the coefficient of $\xi\psi_3$ to unity and parameter $B_3$ is chosen so that the quantity in brackets in Eq. (32) is zero. Parameters $A_3$ and $B_3$ are in Eqs. (33) and (34) to simplify Eq. (31) to give Eq. (35). Note that $A_3$ is greater than zero and $B_3$ is less than zero.

$$A_3 = \left[\frac{\hbar^2(S+a)}{2mU_0}\right]^{\frac{1}{3}} \quad (33)$$

$$B_3 = -(S+a)\frac{E}{U_0} - S \quad (34)$$

$$\frac{d^2\psi_3}{d\xi^2} + \xi\psi_3 = 0 \quad (35)$$

The solution of Eq. (35) is given in Eq. (10) where Ai and Bi are Airy functions [20]. Equation (4) is used with Eq. (36) to obtain Eq. (37) with x as the independent variable. Note that in Region 3 the sign of the argument for the Airy functions is always negative to give a quasi-sinusoidal behavior that corresponds to the energy being greater than the potential. Taking the derivative of Eq. (37) gives Eq. (38) for the derivative of the wavefunction in Region 3.

$$\psi_3(\xi) = C_5 Ai(-\xi) + C_6 Bi(-\xi) \quad (36)$$

$$\psi_3(x) = C_5 Ai\left(\frac{B_3 - x}{A_3}\right) + C_6 Bi\left(\frac{B_3 - x}{A_3}\right) \quad (37)$$

$$\frac{d\psi_3}{dx} = -\frac{C_5}{A_3} Ai'\left(\frac{B_3 - x}{A_3}\right) - \frac{C_6}{A_3} Bi'\left(\frac{B_3 - x}{A_3}\right) \quad (38)$$

The wavefunction and its derivative in Region 1 where -S < x < 0 are given by Eqs. (39), (40), and (41).

$$\psi_1(x) = C_1 e^{-ik_1 x} + C_2 e^{ik_1 x} \quad (39)$$

$$k_1 = \frac{\sqrt{2mE}}{\hbar} \quad (40)$$

$$\frac{d\psi_1}{dx} = -ik_1 C_1 e^{-ik_1 x} + ik_1 C_2 e^{ik_1 x} \quad (41)$$

Following the procedure that was used in Region 3, the wavefunction and its derivative in Region 2 where 0 < x < a are given by Eqs. (42), (43), (44), and (45).

$$\psi_2(x) = C_3 Ai\left(\frac{B_2 - x}{A_2}\right) + C_4 Bi\left(\frac{B_2 - x}{A_2}\right) \quad (42)$$

$$A_2 = \left[\frac{\hbar^2 a}{2mU_0}\right]^{\frac{1}{3}} \quad (43)$$

$$B_2 = \frac{(\phi - E)}{U_0} a \quad (44)$$

$$\frac{d\psi_2}{dx} = -\frac{C_3}{A_2} Ai'\left(\frac{B_2 - x}{A_2}\right) - \frac{C_4}{A_2} Bi'\left(\frac{B_2 - x}{A_2}\right) \quad (45)$$

Next the two boundary conditions, that the wavefunction and its derivative are continuous, are applied at each of the three boundaries. First Eqs. (46) and (47) are obtained at x = 0 between Region 1 and Region 2.

$$C_1 + C_2 - C_3 Ai\left(\frac{B_2}{A_2}\right) - C_4 Bi\left(\frac{B_2}{A_2}\right) = 0 \quad (46)$$

$$ik_1 C_1 - ik_1 C_2 - \frac{C_3}{A_2} Ai'\left(\frac{B_2}{A_2}\right) - \frac{C_4}{A_2} Bi'\left(\frac{B_2}{A_2}\right) = 0 \quad (47)$$

Applying the boundary conditions at x = a, between Region 2 and Region 3 gives Eqs. (48) and (49).

$$C_3 Ai\left(\frac{B_2 - a}{A_2}\right) + C_4 Bi\left(\frac{B_2 - a}{A_2}\right) - C_5 Ai\left(\frac{B_3 - a}{A_3}\right) - C_6 Bi\left(\frac{B_3 - a}{A_3}\right) = 0 \quad (48)$$

$$\frac{C_3}{A_2} Ai'\left(\frac{B_2-a}{A_2}\right) + \frac{C_4}{A_2} Bi'\left(\frac{B_2-a}{A_2}\right) - \frac{C_5}{A_3} Ai'\left(\frac{B_3-a}{A_3}\right) - \frac{C_6}{A_3} Bi'\left(\frac{B_3-a}{A_3}\right) = 0 \quad (49)$$

Applying the boundary conditions at x = -S, between Region 3 and Region 1 gives Eqs. (50) and (51).

$$C_1 e^{ik_1 S} + C_2 e^{-ik_1 S} - C_5 Ai\left(\frac{B_3+S}{A_3}\right) - C_6 Bi\left(\frac{B_3+S}{A_3}\right) = 0 \quad (50)$$

$$ik_1 C_1 e^{ik_1 S} - ik_1 C_2 e^{-ik_1 S} - \frac{C_5}{A_3} Ai'\left(\frac{B_3+S}{A_3}\right) - \frac{C_6}{A_3} Bi'\left(\frac{B_3+S}{A_3}\right) = 0 \quad (51)$$

Notice that Eqs. (46), (47), (48), (49), (50), and (51) form a system of 6 simultaneous equations in the 6 unknown coefficients. We label the determinant for the coefficients "det", where $M_{IJ}$ is the term in the I-th row and J-th column. This determinant has only three non-zero products as shown in Eq. (52).

$$\det = M_{11} M_{22} M_{33} M_{44} M_{55} M_{66} + M_{12} M_{23} M_{34} M_{45} M_{56} M_{61} + M_{13} M_{24} M_{35} M_{46} M_{51} M_{62} = 0 \quad (52)$$

Substituting the expressions for the matrix elements into Eq. (52) and removing a common prefactor gives Eq. (53).

$$Ai\left(\frac{B_2-a}{A_2}\right) Bi'\left(\frac{B_2-a}{A_2}\right) Ai\left(\frac{B_3+S}{A_3}\right) Bi'\left(\frac{B_3+S}{A_3}\right) + Ai\left(\frac{B_2}{A_2}\right) Bi'\left(\frac{B_2}{A_2}\right) Ai\left(\frac{B_3-a}{A_3}\right) Bi'\left(\frac{B_3-a}{A_3}\right)$$

$$+ Ai'\left(\frac{B_2}{A_2}\right) Bi\left(\frac{B_2-a}{A_2}\right) Ai'\left(\frac{B_3-a}{A_3}\right) Bi\left(\frac{B_3+S}{A_3}\right) e^{ik_1 S} = 0 \quad (53)$$

Note that the first two terms in Eq. 53) are real and the third term is complex because of the imaginary exponential. Thus, we require that Eq. (54) be satisfied so that there will be no imaginary component. Thus, there are two solutions given by Eqs. (55 and (56).

$$k_1 S = n\pi \quad (54)$$

For n even:

$$Ai\left(\frac{B_2-a}{A_2}\right) Bi'\left(\frac{B_2-a}{A_2}\right) Ai\left(\frac{B_3+S}{A_3}\right) Bi'\left(\frac{B_3+S}{A_3}\right) + Ai\left(\frac{B_2}{A_2}\right) Bi'\left(\frac{B_2}{A_2}\right) Ai\left(\frac{B_3-a}{A_3}\right) Bi'\left(\frac{B_3-a}{A_3}\right)$$

$$+ Ai'\left(\frac{B_2}{A_2}\right) Bi\left(\frac{B_2-a}{A_2}\right) Ai'\left(\frac{B_3-a}{A_3}\right) Bi\left(\frac{B_3+S}{A_3}\right) = 0 \quad (55)$$

For n odd:

$$Ai\left(\frac{B_2-a}{A_2}\right) Bi'\left(\frac{B_2-a}{A_2}\right) Ai\left(\frac{B_3+S}{A_3}\right) Bi'\left(\frac{B_3+S}{A_3}\right) + Ai\left(\frac{B_2}{A_2}\right) Bi'\left(\frac{B_2}{A_2}\right) Ai\left(\frac{B_3-a}{A_3}\right) Bi'\left(\frac{B_3-a}{A_3}\right)$$

$$- Ai'\left(\frac{B_2}{A_2}\right) Bi\left(\frac{B_2-a}{A_2}\right) Ai'\left(\frac{B_3-a}{A_3}\right) Bi\left(\frac{B_3+S}{A_3}\right) = 0 \quad (56)$$

Equation (54) requires that the length S of the pre-barrier region be an integer multiple of one-half of the De Broglie wavelength as shown in Eq. (57), where the De Broglie wavelength is defined in Eq. (58) with the symbol "p" standing for the momentum of the electron.

$$S = \frac{n\lambda_{dB}}{2} \quad (57)$$

$$\lambda_{dB} \equiv \frac{h}{p} \qquad (58)$$

Standing waves must have a constant potential over a defined spatial extent so we do not see them in the sections of our models that have represent either an ideal resistor or an ideal battery.

## VI. PROCEDURE FOR APPLYING THIS ANALYSIS

Next, we define an algorithm to apply the analysis in the previous section when modeling the operation of a device.

1. Specify the fundamental constants: Planck's constant and the rest mass and charge of the electron.
2. Specify the work function for the cathode.
3, Specify a, the length of the tunneling junction.
4. Specify $V_0$, the applied potential.
5. Specify the integer n.
6. Specify a trial value for the energy E of the electron relative to the fermi level.
7. Calculate $k_1$, the propagation constant in the pre-barrier region using Eq. (40).
8. Calculate S, the pre-barrier length, using Eq. (54).
9. Calculate the parameters $A_3$, $B_3$, $A_2$, and $B_2$ for determining the Airy functions by using Eqs. (33), (34), (43), and (44).
10. Calculate the 12 Airy functions $Ai(B_2/A_2)$, $Ai'(B_2/A_2)$, $Bi'(B_2/A_2)$, $Ai[(B_2-a)/A_2]$, $Bi[(B_2-a)/A_2]$, $Bi'[(B_2-a)/A_2]$, $Ai[(B_3-a)/A_3]$, $Bi[(B_3-a)/A_3]$, $Bi'[(B_3-a)/A_3]$, $Ai[(B_3+S)/A_2]$, $Bi[(B_3+S)/A_2]$, and $Bi'[(B_3+S)/A_2]$.
11. If the integer n is even use Eq. (55) to obtain the error and if n is odd use Eq. (56).
12. Return to step 6 and use another value of E until the iterations are converged.

This procedure will give the n-th eigenvalue for the energy E that corresponds to the specified values of a and $V_0$.

## VII. EXTENSION BY MEANS OF A QUASISTATIC APPROXIMATION

In Section III it was noted that the mechanism studied by Tien and Gordon [8] requires photon processes with superconducting electrodes to provide a time-dependent tunneling current and they do not have a unique solution of the time-dependent Schrödinger equation. Another exact solution of the time-dependent Schrödinger equation that is also based on photon processes was more recently presented by Zhang and Lau [21].

We have generated microwave frequency combs by focusing a mode-locked ultrafast laser on the tunneling junction of a scanning tunneling microscope [5] and our analysis of these measurements suggests that quantum processes are not involved in this application [22]. Others have described photon-assisted processes in analyses [8], [23] and measurements [24], [25], [26]. However, these effects depend on the distance between the electrodes and photon processes would not be expected under the conditions for our measurements [5], [7], [27], [28].

In our first measurements of time-dependent quantum tunneling we connected the primary windings of three audio-frequency transformers in series with a vacuum field emission tube and a DC high-voltage power supply. Two audio-frequency oscillators and an oscilloscope were connected to the secondary windings of the three transformers. We measured tunneling currents at harmonics of the frequencies of the two oscillators as well as components at the mixer frequencies. These results may be understood by considering the nonlinear relationship of the field emission current to the applied voltage. The photon energy was typically 1 peV in these measurements so we do not attribute these effects to photon processes.

To further justify our proposed use of a quasistatic approximation it is also necessary to consider the ratio of the length of the tunneling junction to the wavelength of the laser. The center wavelength for the mode-locked laser that we have used to generate the microwave frequency comb is 800 nm, so the 0.4 nm length of the tunneling junction is only 0.05 percent of the wavelength. Thus, we feel justified to extend the calculations with our models by substituting the time-dependent potential of $U_0 + U_1 \cos(\omega t)$, or simply $U_1 \cos(\omega t)$, for the $U_0$ in the static solutions that we have obtained.

Figure 5 is the equivalent circuit that we will use to make the extension to quasi-static solutions. Note that the time-dependent tunneling current is divided between the load resistance and the shunting capacitance for the tunneling junction which was not considered in our static solutions of the Schrödinger equation.

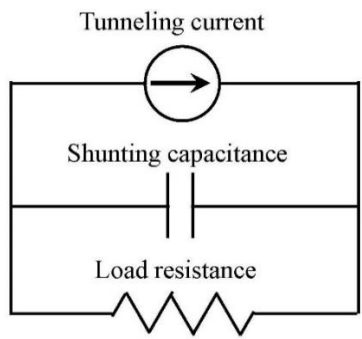

Fig. 5. Equivalent circuit for the quasi-static solutions.

The dependence of the A and B parameters on the potential which is seen in our static solutions causes the tunneling current to have a nonlinear dependence on the applied voltage. Thus, we anticipate that the quasistatic solution for the waveform of the tunneling current which is generated by applying a voltage having an ideal sinusoidal waveform will cause the output to have a finite linewidth. Furthermore, this nonlinearity will cause the superposition of applied voltages at two or more frequencies to have mixing to output currents at frequencies which are the sum and difference of integral multiples of the input frequencies.

## VIII. CONCLUSIONS

It appears necessary to allow for the effects of coherent propagation of the wavefunction through nanoscale circuits because the mean-free path for electrons in metals may be much larger than the size of these circuits. For example, our simulations suggest that standing waves may occur on metallic connections. Methods for quantifying these effects are described. It is also possible to connect the nanoscale circuit to much larger instruments by conventional means. For example, the time-dependent voltage source could be replaced by a nanoscale antenna driven by a full-scale laser. Leads could be used to connect an external voltage source as the "battery" or to monitor the voltage across a nanoscale load resistor.

The mechanisms that are described in this paper raise two separate issues that we suggest to be addressed. Firstly, what effects would they cause in the present devices with or without quantum tunneling? For example, how would the response of a device change if the distribution of energies was changed by varying the temperature or the work function of the cathode in a tunneling junction, or the materials that are used elsewhere in the circuit. It is likely that at present these effects may be interpreted as being a type of noise. Secondly, what are the possibilities for developing new devices that are based on these mechanisms?

**ACKNOWLEDGMENTS**

Dr. Hagmann is grateful to Rolf Landauer and Markus Büttiker who encouraged him to publish his initial work on the numerical modeling of laser-assisted quantum tunneling in 1995, and to Professor Marwan Mousa, at Mu'tah University in Jordan, who made it possible for him to measure laser-assisted field emission during a sabbatical visit to Florida International university from 1999-2000. We are also grateful to Dmitry Yarotski who made it possible to extend these measurements to laser-assisted Scanning Tunneling Microscopy in visits to the Center for Integrated Nanotechnologies (CINT) at Los Alamos National Laboratory from 2008 to 2017. This work was sponsored by the National Science Foundation under Grant 1648811 and the U.S. Department of Energy under Award DE-SC000639.

**APPENDIX**

**1. Piecewise static solution of the Schrödinger equation with a linear potential barrier**

Others have previously solved similar problems using Airy functions [20], [27], [29], [30], but now we present the solution in greater detail because it is used as the first step for the analysis in Section IV of this paper. Figure A1-1 shows the potential energy for a model of quantum tunneling in a static axial electric field. We use the symbol "U" for potential energy to distinguish it from the voltages that are used elsewhere in this paper. A DC electric field of $-U_0/a$ causes the potential to decrease linearly over the length of the barrier. We require $U_0 > 0$ and, $0 < E < \phi - U_0$ as shown in this figure so an electron will propagate classically at the left and right of the barrier and tunnel within the full length of the barrier.

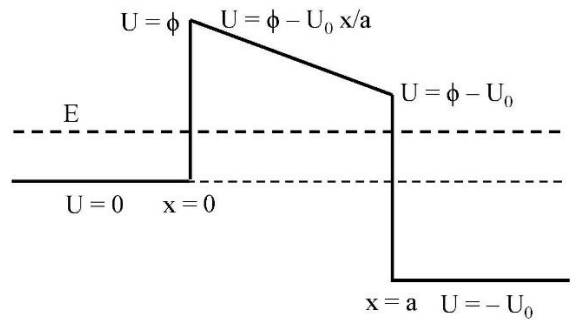

Fig. A1-1. Potential energy for quantum tunneling in a static potential barrier.

With one spatial dimension x, the time-dependent Schrödinger equation is given by Eq. (A1-1). In a static potential this simplifies to Eq. (A1-2) where the wavefunction is given by Eq. (A1-3).

$$\frac{\hbar^2}{2m}\frac{\partial^2 \Psi}{\partial x^2} - U(x,t)\Psi = -i\hbar \frac{\partial \Psi}{\partial t} \qquad (A1-1)$$

$$\frac{\hbar^2}{2m}\frac{d^2\psi}{dx^2} + \left[E - U(x)\right]\psi(x) = 0 \quad (A1-2)$$

$$\Psi(x,t) = \psi(x)e^{-i\frac{Et}{\hbar}} \quad (A1-3)$$

**Solve the Schrödinger equation for x < 0.**

In Region 1 to the left of the barrier, the solution of Eq. (A1-3) is given by Eq. (A1-4) for an incident wave with unit amplitude and a reflected wave with coefficient R:

$$\psi_1 = e^{i\sqrt{2mE}\frac{x}{\hbar}} + \text{R}e^{-i\sqrt{2mE}\frac{x}{\hbar}} \quad (A1-4)$$

**Solve the Schrödinger equation for x > a.**

In Region 3, to the right of the barrier, the wavefunction is given by Eq. (A1-5) with only a transmitted wave having the coefficient T:

$$\psi_3 = Te^{i\sqrt{2m(E+U_0)}\frac{x}{\hbar}} \quad (A1-5)$$

**Solve the Schrödinger equation for 0 < x < a.**

In Region 2, within the barrier, the potential energy is given by Eq. (A1-6) and substituting this into Eq. (A1-2) gives Eq. (A1-7).

$$U_2(x) = \phi - U_0\frac{x}{a} \quad (A1-6)$$

$$\frac{\hbar^2}{2m}\frac{d^2\psi_2}{dx^2} + \left[E - \phi + U_0\frac{x}{a}\right]\psi_2 = 0 \quad (A1-7)$$

A change of variables shown in Eq. (A1-8) is used with Eq. (17) to obtain Eq. (A1-9) where the coefficients A and B have units of meters. Then Eq. (A1-9) is rearranged to obtain Eq. (A1-10).

$$x = A\xi + B \quad (A1-8)$$

$$\frac{\hbar^2}{2mA^2}\frac{d^2\psi_2}{dx^2} + \left[E - \phi + \frac{BU_0}{a}\right]\psi_2 + \frac{AU_0}{a}\xi\psi_2 = 0 \quad (A1-9)$$

$$\frac{d^2\psi_2}{dx^2} + \frac{2mA^2}{\hbar^2}\left[E - \phi + \frac{BU_0}{a}\right]\psi_2 + \frac{2mU_0A^3}{\hbar^2 a}\xi\psi_2 = 0 \quad (A1-10)$$

Next parameter A is chosen by setting the coefficient of the product $\xi\psi_2$ to unity and parameter B is chosen so that the quantity in parentheses in Eq. (A1-10) is zero. Both A and B have units of meters. As noted earlier, $U_0 > 0$ and $E < \phi - U_0$ so that $A > 0$ and $B_1 > a$.

$$A = \left[\frac{\hbar^2 a}{2mU_0}\right]^{\frac{1}{3}} \quad (A1-11)$$

$$B = \left(\frac{\phi - E}{U_0}\right)a \quad (A1-12)$$

Thus, Eq. (A1-10) is simplified to give Eq. (A1-13) which has the solution shown in Eq. (A1-14) where Ai and Bi are Airy functions [20]. The coefficients $C_1$ and $C_2$ have units of inverse meters. Finally, Eq. (A1-8) is used to with Eq. (A1-13) to obtain Eq. (A1-15) where the independent variable is x. Others have used the notation Ai(-x) and Ai(x) in tables for the values of the Airy function to denote the functions for positive or negative arguments where x is a positive number [31]. However, we use the notation that the quantity within the parentheses is simply the argument of the function.

$$\frac{d^2\psi_2}{d\xi^2} + \xi\psi_2 = 0 \tag{A1-13}$$

$$\psi_2(\xi) = C_1 Ai(-\xi) + C_2 Bi(-\xi) \tag{A1-14}$$

$$\psi_2(x) = C_1 Ai\left(\frac{B-x}{A}\right) + C_2 Bi\left(\frac{B-x}{A}\right) \tag{A1-15}$$

**Determine the wavefunction and its derivative at the two boundaries.**

The wavefunction within the barrier has the following values at the ends of the barrier, x = 0 and x = a, where we note that the argument of the Airy functions is greater than zero.

$$\psi_{2(x=0)} = C_1 Ai\left(\frac{B}{A}\right) + C_2 Bi\left(\frac{B}{A}\right) \tag{A1-16}$$

$$\psi_{2(x=a)} = C_1 Ai\left(\frac{B-a}{A}\right) + C_2 Bi\left(\frac{B-a}{A}\right) \tag{A1-17}$$

The derivative of the wavefunction within the barrier is given by Eq. (A18) where Ai' and Bi' are the derivatives of the Ai and Bi functions. Thus, the derivatives at x = 0 and x = a are given in Eqs. (A1-19) and (A1-20):

$$\frac{d\psi_2}{dx} = -\frac{C_1}{A} Ai'\left(\frac{B-x}{A}\right) - \frac{C_2}{A} Bi'\left(\frac{B-x}{A}\right) \tag{A1-18}$$

$$\frac{d\psi_2}{dx}_{(x=0)} = -\frac{C_1}{A} Ai'\left(\frac{B}{A}\right) - \frac{C_2}{A} Bi'\left(\frac{B}{A}\right) \tag{A1-19}$$

$$\frac{d\psi_2}{dx}_{(x=a)} = -\frac{C_1}{A} Ai'\left(\frac{B-a}{A}\right) - \frac{C_2}{A} Bi'\left(\frac{B-a}{A}\right) \tag{A1-20}$$

The following expressions for the wavefunctions to the left and right of their barrier and their derivatives at x = 0 and x = a are obtained by using Eqs. (A1-4) and (A1-5) for the wavefunctions to the left and right of the barrier.

$$\psi_{1(x=0)} = 1 + R \tag{A1-21}$$

$$\frac{d\psi_1}{dx}_{(x=0)} = \frac{i}{\hbar}\sqrt{2mE} - \frac{i}{\hbar}\sqrt{2mE}R \tag{A1-22}$$

$$\psi_{3(x=a)} = Te^{i\sqrt{2m(E+U_0)}\frac{a}{\hbar}} \tag{A1-23}$$

$$\frac{d\psi_3}{dx}_{(x=a)} = \frac{i}{\hbar}\sqrt{2m(E+U_0)}Te^{i\sqrt{2m(E+U_0)}\left(\frac{a}{\hbar}\right)} \tag{A1-24}$$

**Apply the boundary conditions to determine the coefficients.**
1. Using Eqs. (A1-16) and (A1-21) for continuity of the wavefunction at x = 0:

$$C_1 Ai\left(\frac{B}{A}\right) + C_2 Bi\left(\frac{B}{A}\right) - R = 1 \tag{A1-25}$$

2. Using Eqs. (A1-17) and (A1-23) for continuity of the wavefunction at x = a:

$$C_1 Ai\left(\frac{B-a}{A}\right) + C_2 Bi\left(\frac{B-a}{A}\right) - Te^{i\sqrt{2m(E+U_0)}\frac{a}{\hbar}} = 0 \quad (A1-26)$$

3. Using Eqs. (A1-19) and (A1-22) for continuity of the spatial derivative at x = 0:

$$\frac{C_1}{A} Ai'\left(\frac{B}{A}\right) + \frac{C_2}{A} Bi'\left(\frac{B}{A}\right) - \frac{i}{\hbar}\sqrt{2mE}\,R = -\frac{i}{\hbar}\sqrt{2mE} \quad (A1-27)$$

4. Using Eqs. (A1-20) and (A1-24) for continuity of the spatial derivative at x = a:

$$\frac{C_1}{A} Ai'\left(\frac{B-a}{A}\right) + \frac{C_2}{A} Bi'\left(\frac{B-a}{A}\right) + \frac{i}{\hbar}\sqrt{2m(E+U_0)}\,Te^{i\sqrt{2m(E+U_0)}\left(\frac{a}{\hbar}\right)} = 0 \quad (A1-28)$$

To simplify the notation, we define the parameters $P_1$, $P_2$, and $P_3$ in Eqs. (A1-29), (A1-30), and (A1-31).

$$P_1 \equiv \frac{i}{\hbar}\sqrt{2mE} \quad (A1-29)$$

$$P_2 \equiv \frac{i}{\hbar}\sqrt{2m(E+U_0)} \quad (A1-30)$$

$$P_3 = e^{i\sqrt{2m(E+U_0)}\left(\frac{a}{\hbar}\right)} \quad (A1-31)$$

Equations (A1-25), (A1-26), (A1-27), and (A1-28), constitute a system of 4 simultaneous equations in the 4 coefficients $C_1$, $C_2$, R, and T, which has been simplified to give Eqs. (A1-32), (A1-33), (A1-34), and (1-35).

$$C_1 Ai\left(\frac{B}{A}\right) + C_2 Bi\left(\frac{B}{A}\right) - R = 1 \quad (A1-32)$$

$$C_1 Ai\left(\frac{B-a}{A}\right) + C_2 Bi\left(\frac{B-a}{A}\right) - P_3 T = 0 \quad (A1-33)$$

$$C_1 Ai'\left(\frac{B}{A}\right) + C_2 Bi'\left(\frac{B}{A}\right) - AP_1 R = -AP_1 \quad (A1-34)$$

$$C_1 Ai'\left(\frac{B-a}{A}\right) + C_2 Bi'\left(\frac{B-a}{A}\right) - AP_2 P_3 T = 0 \quad (A1-35)$$

Solving this system of four equations with four unknown coefficients gives the following expressions for the coefficients $C_1$, $C_2$, R, and T, with D which is the denominator in each of these four equations.

$$C_1 = \frac{2AP_1}{D}\left[Bi'\left(\frac{B-a}{A}\right) - AP_2 Bi\left(\frac{B-a}{A}\right)\right] \quad (A1-36)$$

$$C_2 = \frac{2AP_1}{D}\left[AP_2 Ai\left(\frac{B-a}{A}\right) - Ai'\left(\frac{B-a}{A}\right)\right] \quad (A1-37)$$

$$R = \frac{2AP_1}{D}\left[Bi'\left(\frac{B-a}{A}\right) - AP_2 Bi\left(\frac{B-a}{A}\right)\right] Ai\left(\frac{B}{A}\right)$$

$$+ \frac{2AP_1}{D}\left[AP_2 Ai\left(\frac{B-a}{A}\right) - Ai'\left(\frac{B-a}{A}\right)\right] Bi\left(\frac{B}{A}\right) - 1 \quad (A1-38)$$

$$T = \frac{2AP_1}{P_3 D}\left[Bi'\left(\frac{B-a}{A}\right) - AP_2 Bi\left(\frac{B-a}{A}\right)\right] Ai\left(\frac{B-a}{A}\right)$$
$$+ \frac{2AP_1}{P_3 D}\left[AP_2 Ai\left(\frac{B-a}{A}\right) - Ai'\left(\frac{B-a}{A}\right)\right] Bi\left(\frac{B-a}{A}\right) \quad (A1-39)$$

$$D = \left[Bi'\left(\frac{B}{A}\right) - AP_1 Bi\left(\frac{B}{A}\right)\right]\left[Ai'\left(\frac{B-a}{A}\right) - AP_2 Ai\left(\frac{B-a}{A}\right)\right]$$
$$- \left[Ai'\left(\frac{B}{A}\right) - AP_1 Ai\left(\frac{B}{A}\right)\right]\left[Bi'\left(\frac{B-a}{A}\right) - AP_2 Bi\left(\frac{B-a}{A}\right)\right] \quad (A1-40)$$

The probability of tunneling is given by Eq. (A1-41), which is simplified to give Eq. (A1-42).

$$TT^* = \left\{\begin{array}{l} \frac{2AP_1}{P_3 D}\left[Bi'\left(\frac{B-a}{A}\right) - AP_2 Bi\left(\frac{B-a}{A}\right)\right] Ai\left(\frac{B-a}{A}\right) \\ + \frac{2AP_1}{P_3 D}\left[AP_2 Ai\left(\frac{B-a}{A}\right) - Ai'\left(\frac{B-a}{A}\right)\right] Bi\left(\frac{B-a}{A}\right) \end{array}\right\}$$
$$\left\{\begin{array}{l} \frac{2AP_1^*}{P_3 D^*}\left[Bi'\left(\frac{B-a}{A}\right) - AP_2^* Bi\left(\frac{B-a}{A}\right)\right] Ai\left(\frac{B-a}{A}\right) \\ + \frac{2AP_1^*}{P_3 D^*}\left[AP_2^* Ai\left(\frac{B-a}{A}\right) - Ai'\left(\frac{B-a}{A}\right)\right] Bi\left(\frac{B-a}{A}\right) \end{array}\right\} \quad (A1-41)$$

$$TT^* = \frac{4A^2}{P_3^2 DD^*}\left[Ai\left(\frac{B-a}{A}\right) Bi'\left(\frac{B-a}{A}\right) - Ai'\left(\frac{B-a}{A}\right) Bi\left(\frac{B-a}{A}\right)\right]^2 \quad (A1-42)$$

## 2. Derivation of the identity used to simplify the summations of Bessel functions

Consider the summation in Eq. (A2-1) which we group to form Eq. (A2-2) and then use Euler's rule to obtain Eq. (A2-3). Equations (A2-4) and (A2-5) are two identities from reference [10], which were verified numerically and used to obtain Eq. (A2-6). Then Eq. (A2-6) was simplified to obtain Eq. (A2-7). Finally, Euler's rule was used to obtain Eq. (A2-8) as an alternative expression for the summation S.

$$S \equiv \sum_{-\infty}^{\infty} J_n(\alpha) e^{-in\beta} \quad (A2-1)$$

$$S = J_0(\alpha) - J_1(\alpha)\left[e^{i\beta} - e^{-i\beta}\right] + J_2(\alpha)\left[e^{2i\beta} + e^{-2i\beta}\right]$$
$$- J_3(\alpha)\left[e^{3i\beta} - e^{-3i\beta}\right] + J_4(\alpha)\left[e^{4i\beta} + e^{-4i\beta}\right]$$
$$- J_5(\alpha)\left[e^{5i\beta} - e^{-5i\beta}\right] + J_6(\alpha)\left[e^{6i\beta} + e^{-6i\beta}\right] - \cdots \quad (A2-2)$$

$$S = J_0(\alpha) - 2iJ_1(\alpha)\sin(\beta) + 2J_2(\alpha)\cos(2\beta)$$
$$- 2iJ_3(\alpha)\sin(3\beta) + 2J_4(\alpha)\cos(4\beta)$$
$$- 2iJ_5(\alpha)\sin(5\beta) + 2J_6(\alpha)\cos(6\beta) - \cdots \quad (A2-3)$$

$$\sum_{k=0}^{\infty} J_{2k+1}(\alpha) \sin\left[(2k+1)\beta\right] \equiv \frac{1}{2}\sin\left[\alpha\sin(\beta)\right] \qquad (A2-4)$$

$$\sum_{k=1}^{\infty} J_{2k}(\alpha) \cos(2k\beta) \equiv \frac{1}{2}\cos\left[\alpha\sin(\beta)\right] - \frac{1}{2}J_0(\alpha) \qquad (A2-5)$$

$$S = J_0(\alpha) + 2\left\{\frac{1}{2}\cos\left[\alpha\sin(\beta)\right] - \frac{1}{2}J_0(\alpha)\right\} - 2i\left\{\frac{1}{2}\sin\left[\alpha\sin(\beta)\right]\right\} \qquad (A2-6)$$

$$S = \cos\left[\alpha\sin(\beta)\right] - i\sin\left[\alpha\sin(\beta)\right] \qquad (A2-7)$$

$$S = e^{-i\alpha\sin(\beta)} \qquad (A2-8)$$